\documentclass[
    ,final            
  ]
  {aipproc}

\layoutstyle{6x9}

\begin{document}

\title{Dynamical collective potential energy landscape: its impact on the competition between fusion and 
quasi-fission in a heavy fusing system}

\classification{25.70.Jj,25.60.Pj,24.10.Pa,24.60.Dr}
\keywords      {Fusion, Quasi-fission, Master equation, Two-center shell model, 
                Super-heavy elements}

\author{Alexis Diaz-Torres}{
  address={Department of Nuclear Physics, Research School of Physical 
  Sciences and Engineering, Australian National University, 
  Canberra, ACT 0200, Australia}
}



\begin{abstract}
 A realistic microscopically-based quantum approach to the competition between fusion and 
quasi-fission in a heavy fusing system is applied to several reactions leading to 
$^{256}$No. Fusion and quasi-fission are described in terms of a diffusion process of nuclear shapes 
through a dynamical collective potential energy landscape which is initially diabatic 
and gradually becomes adiabatic. The microscopic ingredients of the theory are obtained with a 
realistic two-center shell model based on Woods-Saxon potentials. The results indicate that 
(i) the diabatic effects play a very important role in the onset of fusion hindrance for heavy 
systems, and (ii) very asymmetric reactions induced by closed shell nuclei seem to be the best 
suited to synthesize the heaviest compound nuclei.
\end{abstract}

\maketitle


\section{Introduction}

The understanding of the formation mechanism of heavy and super-heavy 
systems in fusion reactions, translated to a problem of diffusion of 
a many-body quantum system on a multi-dimensional collective 
potential energy surface (PES), is still a challenge for present-day theory. 
There is no common viewpoint (see Ref. \cite{Alexis0} for details) 
regarding the modelling of the 
intermediate stage of evolution of the compact nuclear shapes towards 
the compound nucleus (CN) formation in competition with quasi-fission. 
Quasi-fission means re-separation of the dinuclear molecular complex 
before the CN formation. The main motivations for the present work have been 
(i) to reconcile the current conflicting models for CN formation, and (ii) to 
incorporate the \textit{multi-particle quantum nature} of the fusing system, rather than 
assuming a continuous macroscopic fluid. This may explain and predict the dependence 
of fusion (after contact of the nuclei) on the nuclear structure of the interacting 
nuclei.


\section{Theory and numerical results}

\subsubsection{Dynamical collective potential energy surface}

The main feature of the present theory is the concept of 
dynamical collective PES. This PES is obtained with Strutinsky's method 
after (i) solving the two-center problem for fusion microscopically 
\cite{Alexis1}, and (ii) using the idea of the dissipative diabatic 
dynamics suggested by N\"orenberg in Ref. \cite{Noerenberg} 
(see Fig. 1). 

\begin{figure}
  \includegraphics[height=.2\textheight]{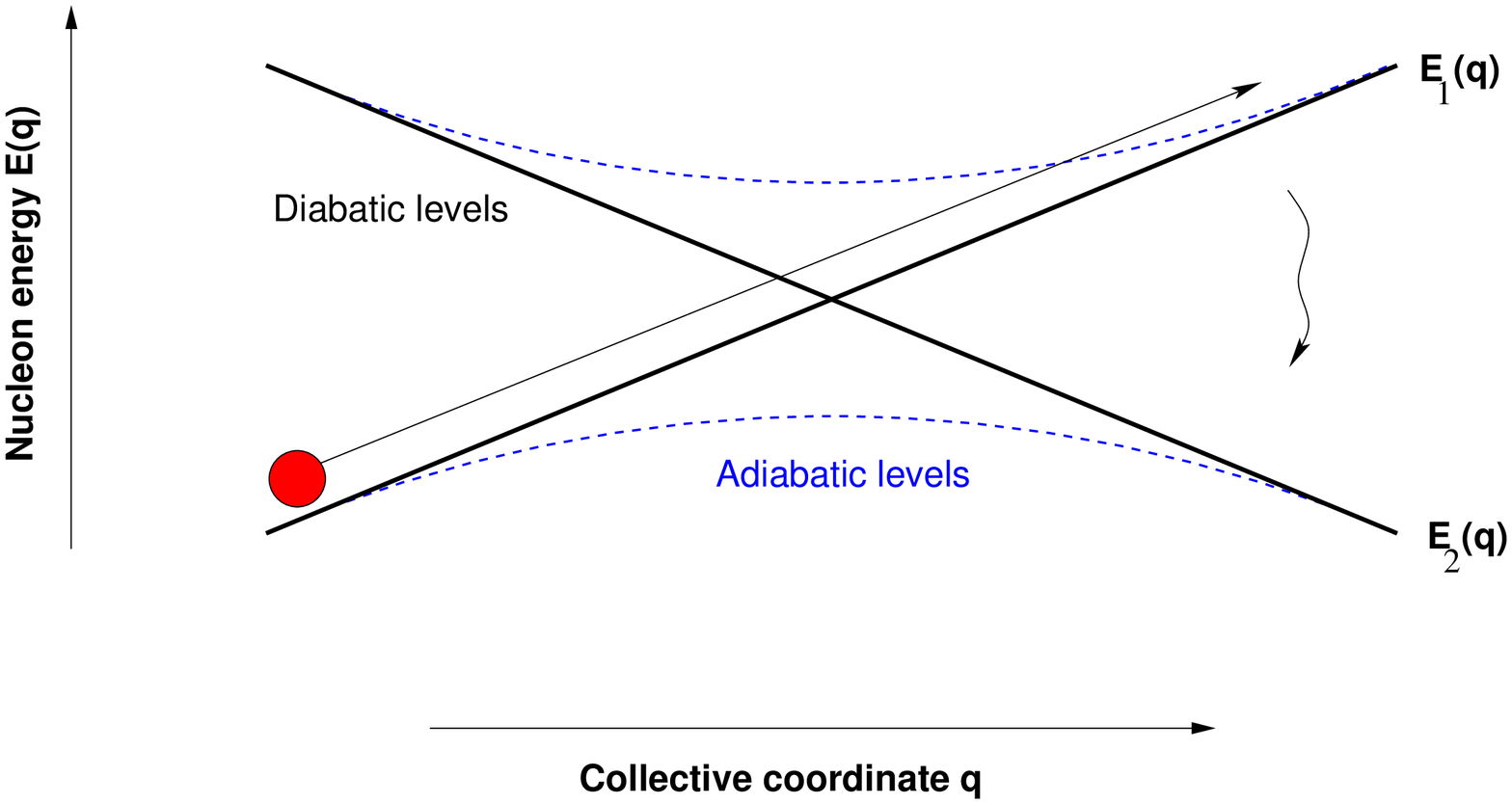}
  \caption{Schematic illustration of the diabatic single-particle 
           motion (solid curves) at an avoided crossing of two adiabatic 
           single-particle levels (dashed curves) with the same quantum 
           numbers. See text for further details.}
\end{figure}


In Fig. 1 a typical avoided crossing of two molecular adiabatic single-particle 
(sp) levels with the same quantum numbers (dashed curve) is shown schematically. 
Many of these pseudo-crossings can occur around the Fermi surface of the fusing system. 
Adiabatic sp orbitals diagonalize the two-center sp Hamiltonian, whereas the diabatic states 
(solid curves) minimize the strong dynamical non-adiabatic coupling 
(induced by the collective kinetic energy operator $\sim \partial /\partial q$) at an avoided 
crossing between two adiabatic sp levels (see Ref. \cite{Alexis1}). 
In the entrance phase of the reaction, the nucleons can follow diabatic levels 
\cite{Noerenberg} instead of remaining in the lowest adiabatic sp orbits. 
This mechanism destroys the Fermi distribution of the sp occupation numbers 
(there is no thermal equilibrium in the system and, strictly speaking, 
the concept of temperature is meaningless). Diabaticity only produces coherent 
particle-hole (ph) excitations that contribute to the collective PES. The 
two-body residual interactions gradually destroy these coherent ph excitations, 
i.e., the system heats up and the sp occupation numbers evolve in time towards a 
Fermi distribution for a finite temperature. The evolution in time of the nuclear shapes 
(a non-equilibrated macroscopic process) occurs in conjunction with the intrinsic thermalization 
of each nuclear shape. The multi-dimensional collective PES, on which the nuclear shapes 
diffuse, is dynamical. This is initially diabatic (a sort of sudden PES) and gradually becomes adiabatic. 

\begin{figure}
  \includegraphics[height=.3\textheight]{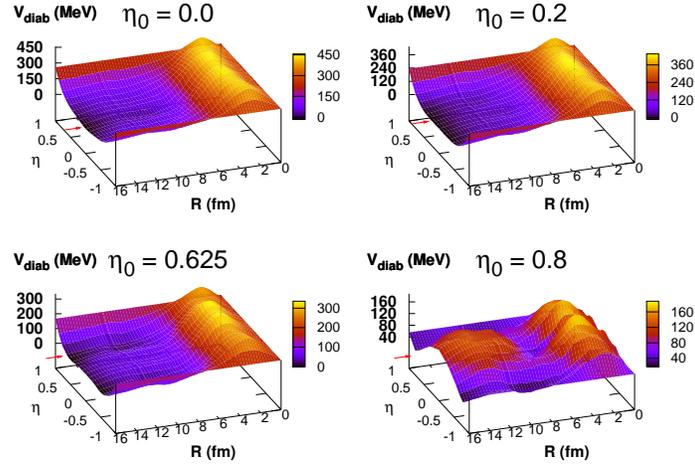}
  \caption{Entrance diabatic PES as a function of the separation between the nuclei R 
           and the mass partition $\eta$ for different entrance channels $\eta_0$ 
           leading to $^{256}$No. See text for further details.}
\end{figure}

\begin{figure}
  \includegraphics[height=.3\textheight]{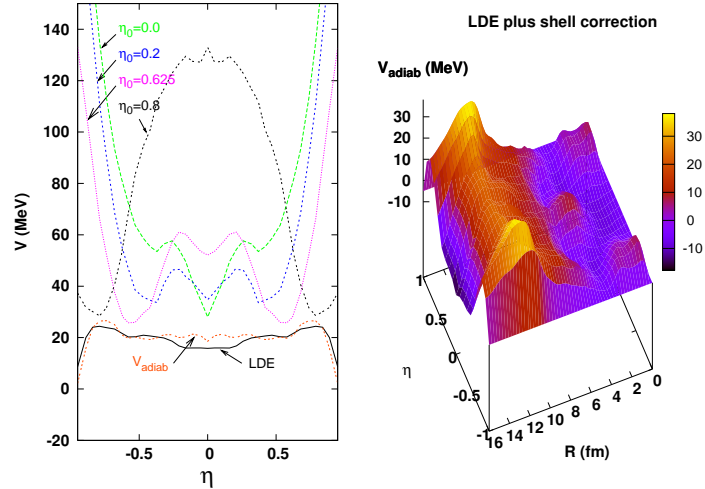}
  \caption{(Left) Entrance driving potentials (curves other than the two lowest ones) 
           for the reactions presented in 
           Fig. 2. The two lowest curves are the asymptotic adiabatic driving potential 
           (red dotted curve) and the liquid drop energy (LDE) only 
           (black solid curve). 
           (Right) Adiabatic PES as a function of R and $\eta$ for the 
           cold system $^{256}$No. See text for further details.}
\end{figure}
 
Fig. 2 shows the entrance diabatic collective PES (liquid drop energy + 
shell corrections + diabatic contribution) as a function of the separation R between 
the nuclei and the mass partition (mass asymmetry $\eta$) for different entrance channels $\eta_0$
(target-projectile combinations) leading to the CN $^{256}$No. This PES reveals a strong repulsive core at small radii R and at large mass asymmetry $\eta$. The initial 
configuration of the sp occupation numbers of the separated interacting nuclei along with the shell structure of the different nuclear shapes determine these PES. 
Fig. 3 (left) shows a cut of these PES along the mass asymmetry coordinate $\eta$ at the contact radius of 
the different fragmentations (entrance driving potentials). The two lowest curves correspond to the adiabatic driving potential V$_{adiab}$ (liquid drop energy + shell corrections, red dotted curve) and the liquid drop energy (LDE) only (black solid curve). The entrance driving potential is raised with respect to the adiabatic potential due to diabatic effects. Please note that there are shell effects in the entrance driving potentials that are not related to the static ground-state shell corrections, but to the diabatic sp motion through the shell structure of the different nuclear shapes. These dynamical shell effects are reflected in the mass yield of the quasi-fission fragments discussed below. 
During the fusion process the entrance diabatic PES (Fig. 2) relaxes to an 
asymptotic adiabatic PES which is less structured than that shown in the right panel in 
Fig. 3 (here the shell corrections are calculated at zero temperature) because the shell corrections decrease at a (local) finite temperature of the nuclear shapes.     
The competition between fusion and quasi-fission is described as a diffusion process of nuclear shapes through this dynamical collective potential energy landscape, caused by quantum and thermal fluctuations. 

 
\subsubsection{Evolution of the compact nuclear shapes}

This scenario is modelled \cite{Alexis0} solving a set of master equations coupled to 
the relaxation equation for the sp occupation numbers. The transition probability rate 
between the nuclear shapes contains quantum and thermal effects on shape fluctuations 
(see Ref. \cite{Alexis0}). 
Owing to the statistical nature of this approach, there is no equation of motion for 
individual collective coordinates \cite{Aritomo,Zagrebaev}. The master equations only describe the evolution in 
time of an ensemble of nuclear shapes (parametrically defined by a set of collective coordinates) that develop following contact of the interacting nuclei. The basic 
macroscopic variable is the nuclear shape, which determines the two-center mean-field 
in which the nucleons are moving. After capture of the nuclei, the motion of the compact 
fusing system is expected to be slow (overdamped), as the initial diabatic collective PES practically absorbs the total incident energy of the system. 

The system of equations has been solved \cite{Alexis0} in a 2D-model using a mesh with respect to the separation between the nuclei R and the mass asymmetry coordinate $\eta$. Each node on the mesh corresponds to a nuclear shape.
The whole mesh can be divided in three regions: (i) region of compact shapes around the near-spherical shape of the CN (\textit{fusion region}), (ii) region of separated fragments beyond the Coulomb barrier (\textit{quasi-fission region}), and (iii) region of intermediate shapes which could lead to fusion or quasi-fission (\textit{competition region}). The probability for CN formation $P_{CN}$ is defined as the population 
of the fusion region, and the quasi-fission probability $P_{QF}$ is described as the population of the quasi-fission region. The time scale for fusion--quasi-fission 
$\tau_{qf}$ is obtained from the condition that following capture the initial (unit) probability of the colliding nuclei occupying the contact configuration (located in the competition region) becomes zero as the probability distribution bifurcates into the fusion and the quasi-fission regions. The mass yield of the quasi-fission fragments is calculated by projecting the quasi-fission probability $P_{QF}$ along the mass asymmetry coordinate $\eta$. It is important to emphasize that our theoretical mass distribution does not include the fission component (decay of the CN into two fragments), but is limited to all binary fragmentations which occur after capture and before the CN formation. This is exactly what we call quasi-fission. 

\begin{figure}
  \includegraphics[height=.3\textheight]{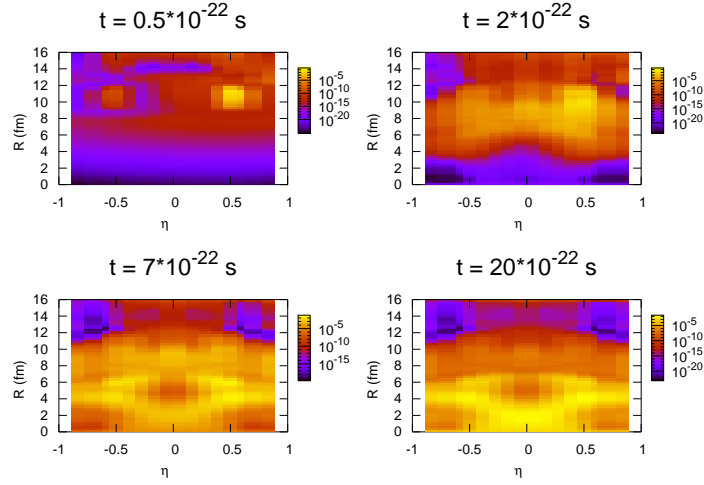}
  \caption{Evolution in time of the probability distribution of the nuclear shapes 
          (dynamical solution of the master equations) for 
          $^{48}$Ca + $^{208}$Pb $\to$ $^{256}$No ($\eta_{0}=0.625$). 
          See text for further details.}
\end{figure}

Fig. 4 shows the evolution in time of the probability distribution of the nuclear shapes 
(solution of the master equations) on the dynamical PES for the reaction 
$^{48}$Ca + $^{208}$Pb $\to$ $^{256}$No. The collision is central and the total incident 
energy in the center of mass (cm) frame is 30 MeV. Further details can be found in Ref. \cite{Alexis0}. It is important to note that the colour scale covers thirty orders of magnitude. In this picture we can see that initially (time up to $2*10^{-22}$ s) the distribution of probability mainly spreads along the nuclear shapes at the contact separation. During this period of time the mass asymmetry coordinate $\eta$ plays the relevant role. Later on, when the diabatic collective PES has completely relaxed to the adiabatic one (relaxation time is about $5*10^{-22}$ s), the maximal values of the probability move to the fused compact shapes (small R). 

\subsubsection{Observables}

Fig. 5 shows the dependence of $P_{CN}$ and the quasi-fission mass yield 
($P_{QF}$ vs. $\eta$) on three crucial 
physical parameters that may control the dynamical evolution of the compact nuclear 
shapes, namely the total incident energy $E_{cm}$ (top panels), the total angular momentum $J$ (middle panels) and the entrance channel mass asymmetry $\eta_{0}$ 
(bottom panels). The quasi-fission time $\tau_{qf}$ remains around $10^{-22}$ s 
(see Ref. \cite{Alexis0}). At the top, we can see in the calculation for a central symmetric collision, that the maximum of $P_{CN}$ is around the capture barrier energy ($30$ MeV) 
and slightly decreases with increasing incident energy. At the middle, we see that $P_{CN}$ for a symmetric collision at $E_{cm}=30$ MeV weakly depends on the angular momentum for the smaller partial waves up to $40 \hbar$. As discussed in detail in Ref. \cite{Alexis0}, the dependence of 
$P_{CN}$ on $E_{cm}$ and $J$ can be explained in terms of the competition between the phase space of the quasi-fission fragments and the fused nuclear shapes. 
In the mass yields for a symmetric entrance channel (top and middle right panels), we can see structures which are related to the structures (valleys) of the entrance diabatic collective PES (see left panel in Fig. 3), while the main peak corresponds to the entrance channel mass asymmetry 
$\eta_{0}=0.0$. In contrast to this reaction, the quasi-fission mass yield for 
$^{48}$Ca + $^{208}$Pb ($\eta_{0}=0.625$ in the bottom right panel) shows a minimum at this entrance mass partition. Here the calculations are for a central collision 
at $E_{cm}=30$ MeV. This minimum is associated with a local maximum for $P_{CN}$ as 
presented in the bottom left panel (see small inserted picture). In this panel, the black solid curve includes all shell effects (shell corrections + diabatic effects). 
In the blue dotted curve, the shell corrections are removed. The red dashed curve in the top left corner of the small figure inserted is without diabatic effects. What this figure tells us is that (i) diabatic effects can tremendously inhibit the fusion of near-symmetric systems, and (ii) remaining ground-state shell corrections to the collective PES can be very important in establishing the $P_{CN}$ value.

\begin{figure}
  \includegraphics[height=.4\textheight]{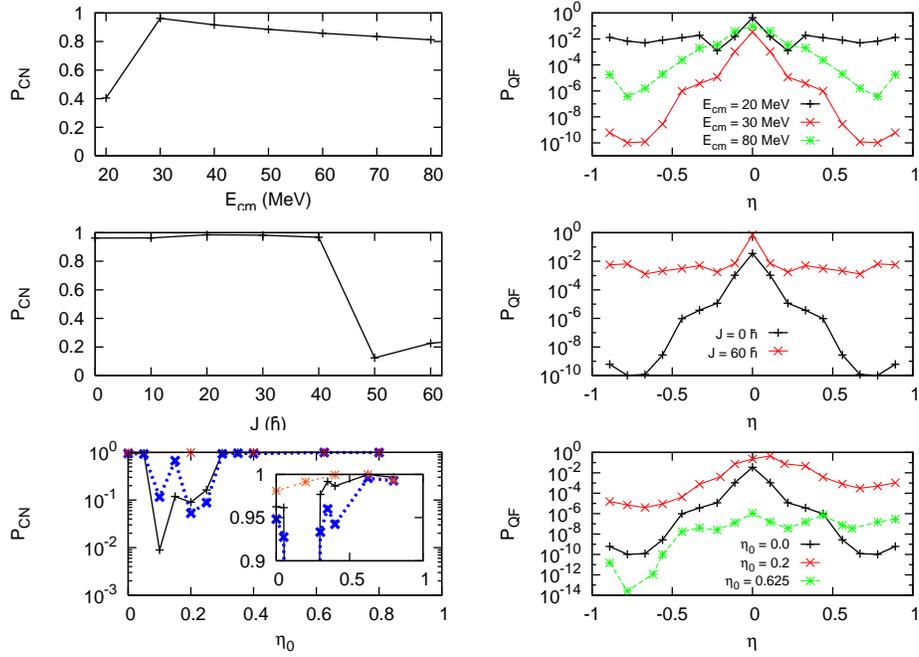}
  \caption{Dependence of $P_{CN}$ and the quasi-fission mass yield 
           ($P_{QF}$ vs. $\eta$) on the 
           total incident energy $E_{cm}$ (top panels), the total angular momentum 
           $J$ (middle panels) and the entrance channel mass asymmetry 
           $\eta_{0}$ (bottom panels). See text for further details.}
\end{figure}




\section{Concluding remarks}

The present preliminary calculations show that (i) the dynamical 
collective PES partially reconciles conflicting aspects of 
current models for CN formation, because both collective 
coordinates mass asymmetry $\eta$ and internuclear distance $R$ 
play a crucial role in fusion ($\eta$ at the beginning of the 
reaction and $R$ towards the end when the diabatic PES has relaxed to 
the adiabatic one), (ii) the diabatic effects and the shell 
corrections are very important in the onset of fusion hindrance 
for heavy systems because they strongly influence the topology of 
the collective PES, and (iii) very asymmetric reactions induced by 
closed shell nuclei seem to be the best suited to form the heaviest 
CN because the diabatic effects are minimized and the contact 
configuration is compact and well inside the capture barrier radius.





\bibliographystyle{aipproc}   

\bibliography{sample}

\begin{thebibliography}{9}

\bibitem{Alexis0}
A.~Diaz-Torres, submitted to \emph{Phys. Rev. C} (2006), 
Arxiv: nucl-th/\textbf{0601042}.

\bibitem{Alexis1}
A.~Diaz-Torres and W.~Scheid, \emph{Nucl. Phys. A} \textbf{757},
  373 (2005).

\bibitem{Noerenberg}
W.~N\"orenberg, \emph{Phys. Lett. B} \textbf{104},
  107 (1981).

\bibitem{Aritomo}
Y.~Aritomo and M.~Ohta, \emph{Nucl. Phys. A} \textbf{744},
  152 (2005), Proceedings of this Conference.

\bibitem{Zagrebaev}
V.I.~Zagrebaev and W.~Greiner, \emph{J. of Phys. G} \textbf{31},
  825 (2005), Proceedings of this Conference.

The author acknowledges the support of an ARC Discovery Grant.

\end{thebibliography}

\IfFileExists{\jobname.bbl}{}
 {\typeout{}
  \typeout{******************************************}
  \typeout{** Please run "bibtex \jobname" to optain}
  \typeout{** the bibliography and then re-run LaTeX}
  \typeout{** twice to fix the references!}
  \typeout{******************************************}
  \typeout{}
 }

\end{document}